\documentclass[prd,showpacs,showkeys]{revtex4}
\usepackage{bm}
\usepackage{graphicx,epsfig}
\usepackage[dvips]{color}
\usepackage{amssymb,amsmath,amsthm}
\begin{document}
\title{Generating perfect fluid spheres in general relativity}
\author{Petarpa Boonserm}
\email{Petarpa.Boonserm@mcs.vuw.ac.nz}
\homepage{http://www.mcs.vuw.ac.nz/research/people/Pertarpa/name}
\affiliation{School of Mathematics, Statistics, and Computer Science, 
Victoria University of Wellington, PO Box 600, Wellington, New Zealand\\}
\author{Matt Visser}
\email{matt.visser@vuw.ac.nz}
\homepage{http://www.mcs.vuw.ac.nz/~visser}
\affiliation{School of Mathematics, Statistics,  and Computer Science, 
Victoria University of Wellington, PO Box 600, Wellington, New Zealand\\}
\author{Silke Weinfurtner}
\email{silke.weinfurtner@vuw.ac.nz}
\homepage{http://www.mcs.vuw.ac.nz/research/people/Silke/name}
\affiliation{School of Mathematics, Statistics,  and Computer Science, 
Victoria University of Wellington, PO Box 600, Wellington, New Zealand\\}
\date{Version 1.0 --- 01 March 2005; 
\LaTeX-ed \today}
\begin{abstract}

\noindent
Ever since Karl Schwarzschild's 1916 discovery of the spacetime geometry describing the interior of a particular idealized general relativistic star --- a static spherically symmetric blob of fluid with position-independent density --- the general relativity community has continued to devote considerable time and energy to understanding the general-relativistic static perfect fluid sphere.
Over the last 90 years a tangle of specific perfect fluid spheres has been discovered,
with most of these specific examples seemingly independent from each other.
To bring some order to this collection, in this article we develop several new transformation theorems that map perfect fluid spheres into perfect fluid spheres.  
These transformation theorems sometimes lead to unexpected connections between previously known perfect fluid spheres, sometimes lead to new previously unknown perfect fluid spheres, and in general can be used to develop a systematic way of classifying the set of all perfect fluid spheres.

\end{abstract}
\maketitle
\newcommand{\diff}[1]{\ensuremath{\mathrm{d}{#1}}}
\newcommand{\dx}[1]{\diff{#1}}
\newcommand{\dr}{\ensuremath{\frac{\mathrm{d}\phantom{r}}{\dx{r}}}}
\newcommand{\eqprime}[1]{\tag{\ref{#1}$^\prime$}}
\newcommand{\dOne}{\ensuremath{\delta_1}}
\newcommand{\dTwo}{\ensuremath{\delta_2}}
\newcommand{\dThr}{\ensuremath{\delta_3}}
\newcommand{\gth}[1]{\textsf{T}_\textsf{#1}}
\newtheorem{theorem}{Theorem}
\newtheorem{lemma}{Lemma}
\newtheorem{corollary}{Corollary}
\newtheorem{definition}{Definition}
\def\d{{\mathrm{d}}}
\def\implies{\Rightarrow}
\def\arctanh{{\mathrm{arctanh}}}
\def\SIM{\triangleq}
\section{Introduction}
\def\txt{\textstyle}
\def\lint{\hbox{\Large $\displaystyle\int$}} 
\def\hint{\hbox{\Huge $\displaystyle\int$}}  
General relativistic perfect fluid spheres, or more precisely general relativistic static perfect fluid spheres, are interesting because they are first approximations to any attempt at building a realistic model for a general relativistic star~\cite{Delgaty,Skea,exact,Martin0}. The central idea is to start solely with spherical symmetry, which implies that in orthonormal components the stress energy tensor takes the form
\begin{equation}
T_{\hat a\hat b} = \left[ \begin{array}{cccc}
\rho&0&0&0\\ 0&p_r&0&0\\ 0&0&p_t&0\\ 0&0&0&p_t \end{array}\right]
\end{equation}
and then use the perfect fluid constraint $p_r=p_t$, making the radial pressure equal to the transverse pressure.
By using the Einstein equations, plus spherical symmetry, the equality  $p_r=p_t$ for the pressures becomes the statement
\begin{equation}
G_{\hat\theta\hat\theta} = G_{\hat r\hat r} = G_{\hat\phi\hat\phi}.
\end{equation}
In terms of the metric components, this leads to an ordinary differential equation [ODE], which then constrains the spacetime geometry, for \emph{any} perfect fluid sphere.

Over the last 90 years, many  ``ad hoc''  approaches to solving this differential equation have been explored, often by picking special coordinate systems, or making simple ansatze for one or other of the metric components~\cite{Buchdahl,Bondi,Wyman,Hojman-et-al}. (For recent overviews see \cite{Delgaty,Skea,exact}.)
The big change over the last several years has been the introduction of ``algorithmic" techniques that permit one to generate large classes perfect fluid spheres in a purely mechanical way~\cite{Rahman,Lake,Martin}. In this article we will extend these algorithmic ideas, by proving several solution-generating theorems of varying levels of complexity.  We shall then explore the formal properties of these solution-generating theorems (which generalize the notion of the Buchdahl transformation) and then will use these theorems to classify some of the previously known exact solutions, and additionally will generate several new previously unknown perfect fluid solutions.


\section{Solution generating theorems}

Start with some static spherically symmetric geometry in Schwarzschild (curvature) coordinates
\begin{equation} \label{line_element_1}
\d s^2 = - \zeta(r)^2 \; \d t^2 + {\d r^2\over B(r)} + r^2 \;\d\Omega^2
\end{equation}
and assume it represents a perfect fluid sphere. That is,
$ G_{\hat\theta\hat\theta} = G_{\hat r\hat r} =G_{\hat\phi\hat\phi} $.
While $G_{\hat\theta\hat\theta} = G_{\hat\phi\hat\phi}$ is always fulfilled due to spherical symmetry,
setting $G_{\hat r\hat r} = G_{\hat\theta\hat\theta}$ supplies us with an ODE
\begin{equation} 
\label{ode_for_B}
[r(r\zeta)']B'+[2r^2\zeta''-2(r\zeta)']B + 2\zeta=0 \, ,
\end{equation}
which reduces the freedom to choose the two functions in equation (\ref{line_element_1})
to one. This equation (\ref{ode_for_B}) is a first-order linear non-homogeneous equation in $B(r)$. Thus
--- once you have chosen a $\zeta(r)$ --- this equation can always be solved for $B(r)$.
Solving for $B(r)$ in terms of $\zeta(r)$ is the basis of~\cite{Lake,Martin}, (and is the basis for Theorem 1 below).
On the other hand, we can also re-group this same equation as
\begin{equation}    
\label{ode_for_zeta}
2 r^2 B \zeta'' + (r^2 B'-2rB) \zeta' +(r B'-2B+2)\zeta=0 \,,
\end{equation}
which is a linear homogeneous second-order ODE for $\zeta(r)$, which will become the basis for Theorem 2 below.
The question we are going to answer in this section is, how to systematically ``deform'' the geometry (\ref{line_element_1}) 
while still retaining the perfect fluid property.
That is, suppose we start with the specific geometry defined by
\begin{equation}
\d s^2 = - \zeta_0(r)^2 \; \d t^2 + {\d r^2\over B_0(r)} + r^2 \d\Omega^2
\end{equation}
and assume it represents a perfect fluid sphere.
We will show how to ``deform'' this solution by applying five different transformation theorems on $\left\{ \zeta_0 , B_0  \right\}$,
such that the outcome still presents a perfect fluid sphere. The outcome of this process will depend on one or more free parameters, and so automatically generates an entire family of perfect fluid spheres of which the original starting point is only one member.
In addition, we analyze what happens if you apply these theorems more than once, iterating them in various ways.

\subsection{Four  theorems}
The first theorem we present is a variant of a result  first explicitly published in \cite{Martin}, though another limited variant of this result  can also be found in~\cite{exact}.
We first re-phrase the theorem of \cite{Martin} in slightly different formalism, and demonstrate an independent way of proving it. Using  our proof it is easy to show that by applying theorem 1 more than once
no further solutions will be obtained, therefore the transformation in theorem 1 is, (in a certain sense to be made precise below), ``idempotent''.

\begin{theorem} 
Suppose $\{ \zeta_0(r), B_0(r) \}$ represents a perfect fluid sphere.
Define
\begin{equation} \label{Theorem1_m_1}
\Delta_0(r)  =
 \left({ \zeta_0(r)\over  \zeta_0(r) + r  \;\zeta'_0(r)}\right)^2 \; r^2 \; 
\exp\left\{ 2 \int {\zeta'_0(r)\over  \zeta_0(r)} \; 
  { \zeta_0(r)- r\; \zeta'_0(r)\over  \zeta_0(r) + r  \;\zeta'_0(r)} \; \d r\right\}.
\end{equation}
Then for all $\lambda$, the geometry defined by holding $\zeta_0(r)$ fixed and
setting
\begin{equation}
\d s^2 = - \zeta_0(r)^2 \; \d t^2 + {\d r^2\over B_0(r)+\lambda\; \Delta_0(r) }
+ r^2 \d\Omega^2
\end{equation}
is also a perfect fluid sphere. That is, the mapping
\begin{equation}
\gth{1}(\lambda): \left\{ \zeta_0 , B_0  \right\} \mapsto 
\left\{ \zeta_0 , B_0 + \lambda\Delta_0(\zeta_0) \right\}
\end{equation}
takes perfect fluid spheres into perfect fluid spheres. Furthermore a second application of the transformation does not yield new information, $\gth{1}$ is ``idempotent'', in the sense that
\begin{equation}
\gth{1}(\lambda_n) \circ \cdots \circ \gth{1}(\lambda_2) \circ \gth{1}(\lambda_1): 
\left\{ \zeta_0 , B_0  \right\} \mapsto 
\left\{ \zeta_0 , B_0 + \left(\sum\nolimits_{i=1}^n \lambda_i \right)\; \Delta_0(\zeta_0) \right\}
\end{equation}
We also note that $\gth{1}$ always has an inverse
\begin{equation}
[\gth{1}(\lambda)]^{-1} = \gth{1}(-\lambda).
\end{equation}
\end{theorem}

\begin{figure}[!ht] 
\centering
\includegraphics[scale = 0.8]{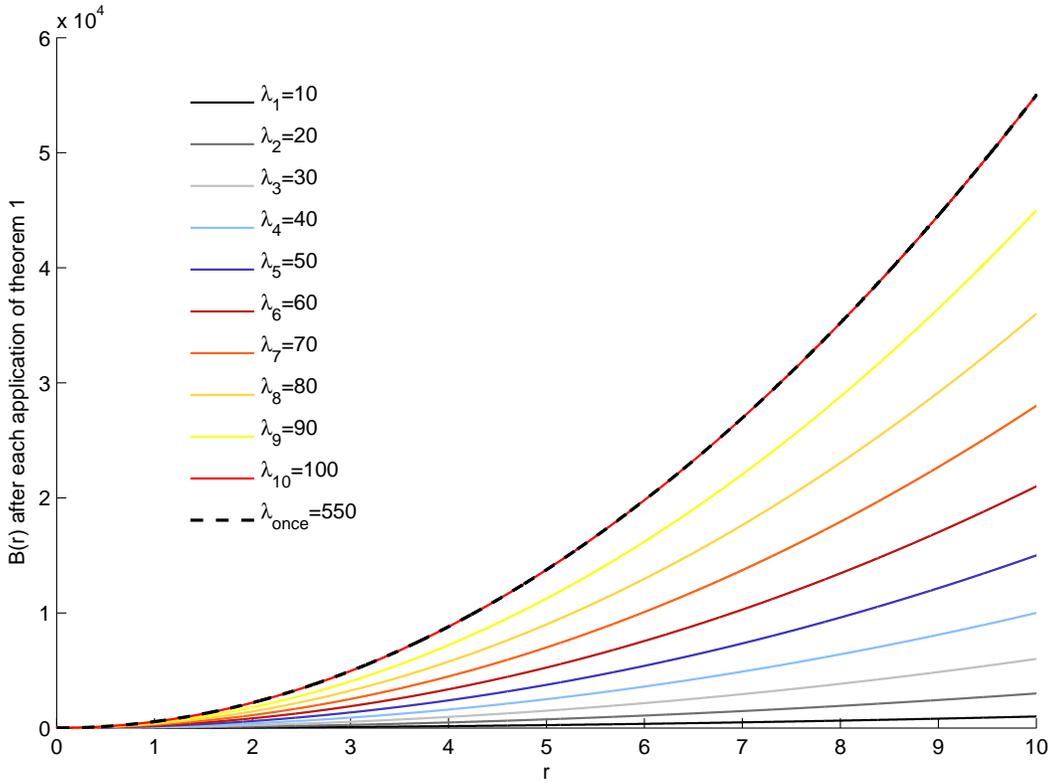}
\caption{\label{Idempotence_Theorem1}
The solid lines show $B(r)$ for  $10$ reapplications of Theorem 1 onto the Minkowski metric.
The dashed line corresponds to a single application with the specific choice $\lambda_{\mathrm{once}}=\sum \lambda_i$. 
It can be seen that $10$ applications of Theorem 1 can be expressed by one application.}
\end{figure}

\begin{proof}[Proof for Theorem 1]
Assume that $\left\{ \zeta_{0}(r),B_{0}(r) \right\}$ is a solution for equation (\ref{ode_for_B}).
Under what conditions does $\left\{ \zeta_{0}(r),B_1(r) \right\}$  also satisfy equation (\ref{ode_for_B})?
Without loss of generality, we write
\begin{equation}
B_1(r)=B_{0}(r) + \lambda\;\Delta_{0}(r) \, .
\end{equation}
Equation ({\ref{ode_for_B}}) can now be used
to determine $\Delta_{0}(r)$.
That ordinary \emph{inhomogeneous} first-order differential equation in $B$ now
simplifies to
\begin{equation}
\label{ode_th1}
\left[ r (r \zeta_0 )' \right] \Delta_{0}'
+ \left[ 2 r^2 \zeta_{0}'' - 2 (r \zeta_{0})' \right] \Delta_{0} = 0   \, ,
\end{equation}
which is an ordinary \emph{homogeneous} first-order differential equation in $\Delta_{0}$.
A straightforward calculation, including an integration by parts, leads to
\begin{equation}
\Delta_0(r) = \frac{r^2 }{\left[ (r \zeta_{0})' \right]^2}  \; \exp\left\{{\int{\frac{4 \zeta_{0}'}{(r \zeta_{0})'} dr } }\right\} \, .
\end{equation}
Adding and subtracting 
$\pm 2 (r \zeta_{0}'^2)/(\zeta_{0} (r \zeta_{0})')$ to the argument under the integral leads to
\begin{equation}
\Delta_{0}
= \left({ \zeta_0(r)\over  \zeta_0(r) + r  \;\zeta'_0(r)}\right)^2 \; r^2 \; 
\exp\left\{ 2 \int {\zeta'_0(r)\over  \zeta_0(r)} \; 
  { \zeta_0(r)- r\; \zeta'_0(r)\over  \zeta_0(r) + r  \;\zeta'_0(r)} \; \d r\right\} \, ,
\end{equation}
as advertised.
If we apply this transformation a second time we obtain no additional information. To see this, consider the sequence
\begin{equation}
\{\zeta_0,B_0\} \mapsto \{\zeta_0,B_1\} \mapsto \{\zeta_0,B_2\} \dots
\end{equation}
But at the second step, since $\zeta_0$ has not changed,  we have $\Delta_1(r)=\Delta_0(r)$. More generally, at all subsequent steps, $\Delta_i(r)=\Delta_0(r)$. We can write this as
\begin{equation}
\prod_{i=1}^n \gth{1}(\lambda_i) = \gth{1}\left(\sum_{i=1}^n \lambda_i \right).
\end{equation}
or in the more suggestive form
\begin{equation}
\prod_{i=1}^n \gth{1}  \SIM \gth{1}
\end{equation}
where the symbol $\SIM$ indicates ``\emph{equality up to relabelling of the parameters}''. 
That is, transformation $\gth{1}$ is ``idempotent'' up to relabelling of the parameters. (See figure \ref{Idempotence_Theorem1}.)
\end{proof}

A version of Theorem 1  can also be found in~\cite{exact}. 
Specifically, after several manipulations, changes of notation, and a change of coordinate system, the transformation exhibited in equation (16.11) of~\cite{exact} can be  converted into the transformation $\gth{1}$ of Theorem 1 above.

Applying theorem 1 to a fixed $\{\zeta_0,B_0\}$ generates a one-dimensional space of perfect fluid spheres, which leads to the corollary below:
\begin{corollary}
Let $\{\zeta_0,B_a\}$ and  $\{\zeta_0,B_b\}$ both represent perfect fluid spheres, then for all $p$
\begin{equation}
\left\{ \zeta_0, p B_a + (1-p) B_b\right\}
\end{equation}
is also a perfect fluid sphere, furthermore all perfect fluid spheres for a fixed $\zeta_0$ can be written in this form.
\end{corollary}

\begin{proof}[Proof]
The result is automatic once you note that for fixed $\zeta_0$ the ODE for $B$ is linear, (though not homogeneous, which is why the two coefficients $p$ and $1-p$ are chosen to add up to 1) .
\end{proof}

We defer extensive discussion of the application of this theorem and its corollary until section \ref{S:classify}, where we use this and our other generating theorems as a basis for classifying perfect fluid spheres. At this stage we mention, only as a very simple example, that $\gth{1}$ applied to Minkowski space results in the geometry of the Einstein static universe.

\begin{theorem} 
Let $\{\zeta_0,B_0\}$ describe a perfect fluid sphere.
Define
\begin{equation}
Z_0(r) = \sigma +\epsilon \lint {r \; \d r\over  \zeta_0(r)^2\; \sqrt{B_0(r)} }.
\end{equation}
Then for all $\sigma$ and $\epsilon$, the geometry defined by holding $B_0(r)$ fixed and
setting
\begin{equation}
\d s^2 = - \zeta_0(r)^2 \; Z_0(r)^2
\; \d t^2 + {\d r^2\over B_0(r)} + r^2 \d\Omega^2
\end{equation}
is also a perfect fluid sphere.
That is, the mapping
\begin{equation}
\gth{2}(\sigma,\epsilon): \left\{ \zeta_0, B_0  \right\} \mapsto \left\{ \zeta_0 \; Z_0(\zeta_0,B_0), B_0 \right\}
\end{equation}
takes perfect fluid spheres into perfect fluid spheres. Furthermore a second application of the transformation does not yield new information, $\gth{2}$ is ``idempotent'' in the sense that
\begin{equation}
\gth{2}(\sigma_n,\epsilon_n) \circ \cdots \circ  \gth{2}(\sigma_3,\epsilon_3) \circ  \gth{2}(\sigma_2,\epsilon_2) \circ  \gth{2}(\sigma_1,\epsilon_1)  = 
\gth{2}(\sigma_n\dots \sigma_3\sigma_2\sigma_1, \; \epsilon_{n\dots321}),
\end{equation}
where
\begin{equation}
\epsilon_{n\dots321} = (\epsilon_1\sigma_2\sigma_3 \cdots \sigma_n)
+ (\sigma_1^{-1} \epsilon_2 \sigma_3 \cdots \sigma_n)
+ (\sigma_1^{-1}\sigma_2^{-1} \epsilon_3 \cdots \sigma_n)
+ \cdots 
+ (\sigma_1^{-1} \sigma_2^{-1} \sigma_3^{-1} \cdots \epsilon_n).
\end{equation}
Furthermore,  theorem 2 is invertible (as long as $\sigma\neq0$):
\begin{equation}
\left[\gth{2}(\sigma,\epsilon)\right]^{-1} = \gth{2}(1/\sigma,-\epsilon).
\end{equation}

\end{theorem}

\begin{proof}[Proof for Theorem 2]
The proof of theorem 2 is based on the technique of ``reduction in order''. 
Assuming that $\left\{ \zeta_{0}(r),B_{0}(r) \right\}$  solves equation (\ref{ode_for_zeta}), 
write 
\begin{equation}
\zeta_1(r)=\zeta_{0}(r) \; Z_0(r) \, .
\end{equation}
and demand that $\left\{ \zeta_{1}(r),B_{0}(r) \right\}$ also solves equation (\ref{ode_for_zeta}).
We find
\begin{equation}
\label{ode_th2}
(r^2 \zeta_0 B_0' + 4 r^2 \zeta_0' B_0 - 2 r \zeta_0 B_0) Z_0' + (2 r^2 \zeta_0 B_0) Z_0''  =0 \, ,
\end{equation}
which is
an ordinary homogeneous second-order differential equation, depending only on $Z_0'$
and $Z_0''$. (So it can be viewed as a first-order homogeneous differential equation in $Z'$, which is solvable.)
Separating the unknown variable to one side,
\begin{equation} \label{de_for_zetaprime}
\frac{Z_0''}{Z_0'}=-\frac{1}{2} \frac{B_0'}{B_0} - 2 \frac{\zeta_0'}{\zeta_0} + \frac{1}{r} \, .
\end{equation}
Re-write $Z_0''/ Z_0' = \d\ln(Z_0')/\d t$, and integrate twice over both sides of
equation (\ref{de_for_zetaprime}), to obtain
\begin{equation} \label{eq_for_zeta_1}
Z_0=\sigma + \epsilon \int{\frac{r \, dr}{\zeta_0(r)^2 \;\sqrt{B_0(r)} }}  \, ,
\end{equation} 
depending on the old solution $\left\{ \zeta_0 (r) , B_0 (r)  \right\}$, and two
arbitrary integration constants $\sigma$ and $\epsilon$.  

To see that the transformation $\gth{2}$ defined in Theorem 2 is ``idempotent'' we first show
\begin{equation}
\gth{2} \circ\gth{2} \SIM \gth{2},
\end{equation}
and then iterate.
The precise two-step composition rule is
\begin{equation}
\gth{2}(\sigma_2,\epsilon_2)\circ\gth{2}(\sigma_1,\epsilon_1) = 
\gth{2}\left(\sigma_2\sigma_1, \;\epsilon_1\sigma_2+{\epsilon_2\over\sigma_1} \right).
\end{equation}
To see ``idempotence'', note that for fixed $B_0(r)$ equation (\ref{ode_for_zeta}) has a solution space that is exactly two dimensional. Since the first application of  $\gth{2}$  takes any specific solution and maps it into the full two-dimensional solution space, any subsequent application of  $\gth{2}$ can do no more than move one around inside this two dimensional solution space --- physically this corresponds to a relabelling of parameters describing the perfect fluid metric you are dealing with, not the generation of new solutions.
To be more explicit about this note that at step one
\begin{equation}
\zeta_0 \to \zeta_1 = \zeta_0 \; \left\{
\sigma_1 +\epsilon_1 \lint {r \; \d r\over  \zeta_0(r)^2\; \sqrt{B_0(r) }}
\right\},
\end{equation}
while at the second step
\begin{equation}
\zeta_1 \to \zeta_2 = \zeta_1 \; \left\{
\sigma_2 +\epsilon_2 \lint {r \; \d r\over  \zeta_1(r)^2\; \sqrt{B_0(r) }}
\right\}.
\end{equation}
That is:
\begin{equation}
\zeta_2 = \zeta_0 \;  \left\{
\sigma_1 +\epsilon_1 \lint {r \; \d r\over  \zeta_0(r)^2\; \sqrt{B_0(r)}} 
\right\}
\;
\left\{
\sigma_2 +\epsilon_2 \lint {r \; \d r\over  \zeta_0(r)^2\; \sqrt{B_0(r)} \;  \;
[
\sigma_1 +\epsilon_1 \int {r \; \d r /( \zeta_0(r)^2\; \sqrt{B_0(r) }}
]^2  }
\right\}.
\end{equation}
But this can be rewritten as
\begin{equation}
\zeta_2 = \zeta_0 \; Z_0 \; 
\left\{ \sigma_2 + {\epsilon_2\over\epsilon_1} \lint { \d Z_0\over Z_0^2 } \right\}
=
\zeta_0 \; Z_0 \; 
\left\{ \sigma_2 - {\epsilon_2\over\epsilon_1} \int { \d \left({1\over Z_0}\right) } \right\} 
= 
\zeta_0 \; Z_0 \; 
\left\{ \sigma_2 - {\epsilon_2\over\epsilon_1}  \left[{1\over Z_0 }-{1\over \sigma_1}\right] \right\}.
\end{equation}
Therefore
\begin{equation}
\zeta_2
=
\zeta_0 \;  \left\{  - {\epsilon_2\over\epsilon_1}  + 
\left[\sigma_2+{\epsilon_2\over\epsilon_1} \;{1\over \sigma_1}\right] \; Z_0 \right\}.
\end{equation}
That is
\begin{equation}
Z_1 =  - {\epsilon_2\over\epsilon_1}  
+ \left[\sigma_2+{\epsilon_2\over\epsilon_1} \;{1\over \sigma_1}\right]  \; Z_0,
\end{equation}
from which the composition law 
\begin{equation}
\gth{2}(\sigma_2,\epsilon_2)\circ\gth{2}(\sigma_1,\epsilon_1) = 
\gth{2}\left(\sigma_2\sigma_1, \;\epsilon_1\sigma_2+{\epsilon_2\over\sigma_1} \right)
\end{equation}
follows immediately. (Note that the composition law for $\gth{2}$ is actually a statement about applying reduction of order to second-order ODEs, it is not specifically a statement about perfect fluid spheres, though that is how we will use it in the current article.)
The general composition law then follows by induction. To find the inverse transformation we choose $\sigma_2=1/\sigma_1$ and $\epsilon_1=-\epsilon_2$, for which
\begin{equation}
\label{E:composition2}
\gth{2}(1/\sigma_1,-\epsilon_1)\circ\gth{2}(\sigma_1,\epsilon_1)= 
\gth{2}\left(1, \;0\right) =  \mathbf{I}.
\end{equation}
\end{proof}

\begin{figure}[!ht]
\centering
\includegraphics[scale = 0.65]{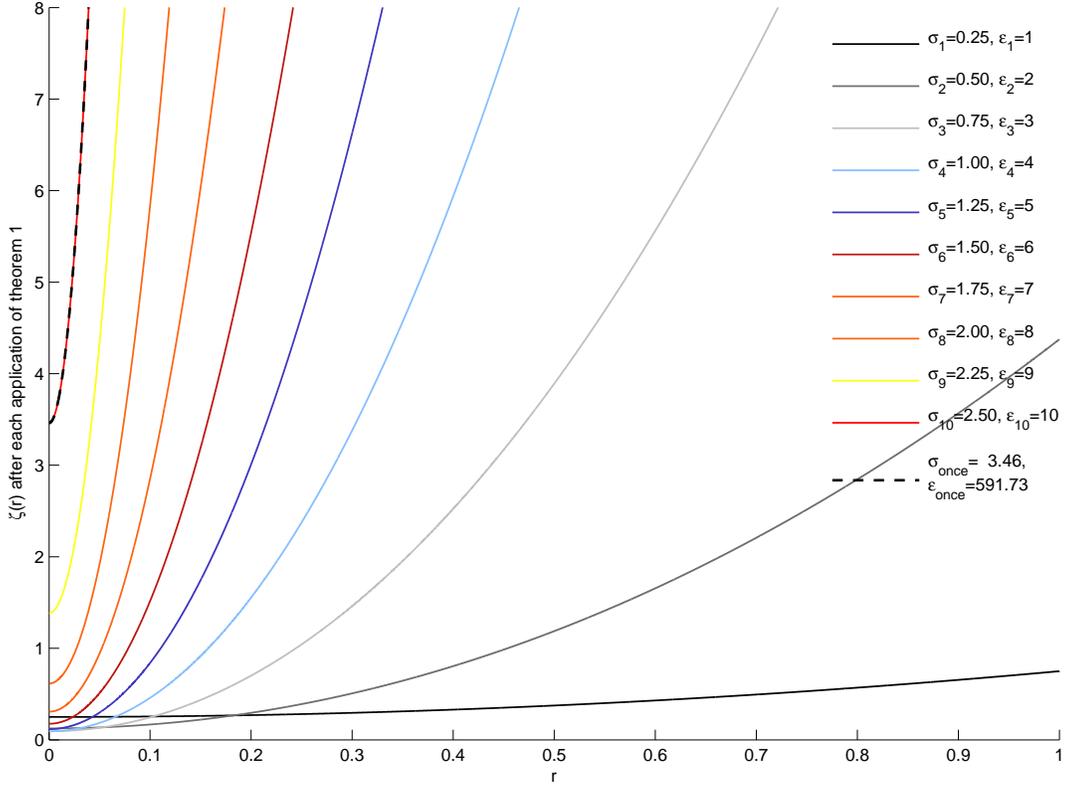}
\caption{\label{Idempotence_Theorem2}
The solid lines show $\zeta(r)$ for  $10$ reapplications of Theorem 2 onto the Minkowski metric.
The dashed line corresponds to a single application with the specific choice for $\sigma_{\mathrm{once}}$ and $\epsilon_{\mathrm{once}}$ determined by the composition law of Theorem 2.
It can be seen that $10$ applications of Theorem 2 can be re-expressed by a single application.}
\end{figure}

Comment: As other special cases of the composition law we also mention the results that
\begin{equation}
\prod_{i=1}^n \gth{2}(1,\epsilon_i)= \gth{2}\left(1,\sum_{i=1}^n \epsilon_i \right),
\end{equation}
and
\begin{equation}
\gth{2}(\sigma,\epsilon)^n= 
\gth{2}\left(\sigma^n,\epsilon\;
\left[\sigma^{n-1}+\sigma^ {n-3} \cdots +\sigma^{-(n-3)}+\sigma^{-(n-1)}\right] 
\right).
\end{equation}
Now as long as $\sigma>1$ then for sufficiently large $n$ we see
\begin{equation}
\gth{2}(\sigma,\epsilon)^n \approx
\gth{2}\left(\sigma^n, \sigma^{n-1} \epsilon \right) = \sigma^{n-1} \; \gth{2}(\sigma,\epsilon) \SIM \gth{2}(\sigma,\epsilon),
\end{equation}
where at the last step we have used the fact that the overall multiplicative factor $\sigma^{n-1}$ can simply be reabsorbed into a redefinition of the time coordinate. Because of this result, we see that for fixed $\sigma>1$ (and fixed but arbitrary $\epsilon$) repeated numerical applications of $\gth{2}(\sigma,\epsilon)$ will have a well-defined limit. In figure \ref{Idempotence_Theorem2} we have tested the composition law numerically.

A strictly limited version of theorem 2,  with little comment on its importance, can be found in~\cite{exact}. 
Specifically, after several manipulations, changes of notation, and a change of coordinate system, the transformation exhibited in equation (16.12) of~\cite{exact} can be seen to be equivalent to the sub-case $\sigma=0$, $\epsilon=1$ of transformation $\gth{2}$ above.

For some purposes it is more useful to rephrase theorem 2 as below:
\begin{corollary}
Let $\{\zeta_a,B_0\}$ and  $\{\zeta_b,B_0\}$ both represent perfect fluid spheres, then for all $p$ and $q$
\begin{equation}
\left\{ p \;\zeta_a + q \;\zeta_b, B_0\right\}
\end{equation}
is also a perfect fluid sphere. Furthermore, for fixed $B_0$ all perfect fluid spheres can be written in this form.
\end{corollary}

\begin{proof}[Proof]
The result is automatic once you note that for fixed $B_0$ the ODE for $\zeta$ is second-order linear and homogeneous.
\end{proof}

We again defer extensive discussion of the application of this theorem and its corollary until section \ref{S:classify}, at this stage mentioning only as a very simple examples that $\gth{2}$ applied to either the Einstein static universe or the anti-de Sitter universe results in Schwarzschild's stellar solution (position-independent density). Similarly, corollary 2 applied to a combination of the Einstein static universe and anti-de Sitter space is another way of obtaining Schwarzschild's stellar solution.

Having now found the first and second generating theorems makes it useful to define two new theorems by composing them.
Take a perfect fluid sphere solution $\left\{ \zeta_0, B_0  \right\}$. Applying theorem 1 onto it gives
us a new perfect fluid sphere $\left\{ \zeta_0, B_1 \right\}$. The new $B_1$ is given in
equation (\ref{Theorem1_m_1}). If we now continue by applying theorem 2, again we get a
new solution $\{ \tilde\zeta, B_1 \}$, where $\tilde{\zeta}$ now depends on the new $B_1$.
All together we can consider this as a single process, by introducing the following theorem: \\

\begin{theorem}
If $\{\zeta_0,B_0\}$ denotes a perfect fluid sphere, then for all $\sigma$,  $\epsilon$, and $\lambda$, the three-parameter geometry defined by
\begin{equation}
\d s^2 = - \zeta_0(r)^2 
\left\{
\sigma+\epsilon \lint {r \; \d r\over  \zeta_0(r)^2\; \sqrt{B_0(r)+\lambda \;\Delta_0(r)}} 
\right\}^2
\; \d t^2 + {\d r^2\over B_0(r) +\lambda \Delta_0(r)   }
+ r^2 \d\Omega^2
\end{equation}
is also a perfect fluid sphere, where $\Delta_0$ is 
\begin{equation}
\Delta_0(r) =
\left({ \zeta_0(r)\over  \zeta_0(r) + r  \;\zeta'_0(r)}\right)^2 \; r^2 \; 
\exp\left\{ 2 \int {\zeta'_0(r)\over  \zeta_0(r)} \; 
  { \zeta_0(r)- r\; \zeta'_0(r)\over  \zeta_0(r) + r  \;\zeta'_0(r)} \; \d r\right\} \, .
\end{equation}
That is
\begin{equation}
\gth{3} = \gth{2}\circ\gth{1}: \{\zeta_0,B_0\} \mapsto \{\zeta_0,B_0+\lambda\;\Delta_0(\zeta_0)\} 
\mapsto \{\zeta_0 \; Z_0(\zeta_0,B_0+\lambda\Delta_0(\zeta_0)),B_0+\lambda\Delta_0(\zeta_0)\}. 
\end{equation}
\end{theorem}

Now, instead of starting with theorem 1 we could first apply theorem 2 on $\left\{ \zeta_0, B_0  \right\}$.
This gives us a new perfect fluid sphere $\left\{ \zeta_1, B_0  \right\}$, 
where $\zeta_1=\zeta_0 \; Z_0(\zeta_0,B_0)$ is given by equation (\ref{eq_for_zeta_1}). Continuing with
theorem 1 leads to $\{ \zeta_1, \tilde{B}  \}$, where $\tilde{B}$ depends on the new
$\zeta_1$. Again, we can consider this as a single process, by introducing the following theorem:

\begin{theorem}
If $\{\zeta_0,B_0\}$ denotes a perfect fluid sphere, then for all $\sigma$,  $\epsilon$, and $\lambda$, the three-parameter geometry defined by
\begin{equation}
\d s^2 = - \zeta_0(r)^2 
\left\{
\sigma+\epsilon \lint {r \; \d r\over  \zeta_0(r)^2\; \sqrt{ B_0(r)   }} 
\right\}^2
\; \d t^2 + {\d r^2\over B_0(r) + \lambda \Delta_0(\zeta_1,r)  }
+ r^2 \d\Omega^2
\end{equation}
is also a perfect fluid sphere, where $\Delta_0(\zeta_1,r)$ is defined as
\begin{equation}
\Delta_0(\zeta_1,r)=
\left({ \zeta_1(r)\over  \zeta_1(r) + r  \;\zeta_1'(r)}\right)^2 \; r^2 \; 
\exp\left\{ 2 \int {\zeta_1'(r)\over  \zeta_1(r)} \; 
  { \zeta_1(r)- r\; \zeta_1'(r)\over  \zeta_1(r) + r  \;\zeta'_1(r)} \; \d r\right\} \, ,
\end{equation}
depending on $\zeta_1 = \zeta_0\;  Z_0$, and where as before
\begin{equation}
Z_0(r) = 
\sigma+\epsilon \lint {r \; \d r\over  \zeta_0(r)^2\; \sqrt{ B_0(r) }}  \, .
\end{equation}
That is
\begin{equation}
\gth{4} = \gth{1}\circ\gth{2}: \{\zeta_0,B_0\} \mapsto \{\zeta_0 \;Z_0(\zeta_0,B_0),B_0\} 
\mapsto \{\zeta_0 \; Z_0(\zeta_0,B_0),B_0+\lambda \Delta_0(\zeta_0 \; Z_0(\zeta_0,B_0))\}.
\end{equation}

\end{theorem}
Theorem 3 and theorem 4 are in general distinct, which can be traced back to the fact that theorem 1 and theorem 2 do not in general commute. Furthermore theorem 3 and theorem 4 are in general not idempotent
\begin{equation}
\gth{3} \not\SIM \gth{4}; \qquad \gth{3} \circ \gth{3} \not\SIM \gth{3}; \qquad  \gth{4} \circ \gth{4} \not\SIM \gth{4}.
\end{equation}
The best way to verify this is to try a few specific examples. There may be some specific and isolated special metrics for which theorem 3 and theorem 4 happen to be degenerate, or idempotent, and finding such metrics is important for our classification programme.

\subsection{Formal properties of the generating theorems}
The solution generating theorems we have developed interact in a number of interesting ways and exhibit numerous formal properties that will be useful when classifying generic perfect fluid spheres.
To start with, theorem 3 $(\gth{3})$ and theorem 4 $(\gth{4})$ can be expressed in terms of theorem 1 $(\gth{1})$ 
and theorem 2 $(\gth{2})$:
\begin{equation}
\begin{split}
\gth{3} (g) :=& \left( \gth{2} \circ \gth{1} \right) g = \gth{2}\left(\gth{1}(g) \right) \, ; \\
\gth{4} (g) :=& \left( \gth{1} \circ \gth{2} \right) g = \gth{1}\left(\gth{2}(g) \right) \, .\\
\end{split} 
\end{equation}
where $g$ is a metric representing a perfect fluid sphere. 
Having this in mind, some of the new solutions generated by starting from some specific solution can be identified. For example:
\begin{equation}
\gth{4} \circ \gth{1}
\equiv \gth{1} \circ \gth{2} \circ \gth{1}
 \equiv \gth{1} \circ \gth{3}  \, ,
\end{equation}
or
\begin{equation}
\gth{3} \circ \gth{3}
\equiv \gth{2} \circ \gth{1} \circ \gth{2} \circ \gth{1}
 \equiv \gth{2} \circ \gth{4} \circ \gth{1}  \, .
\end{equation}
The idempotence of $\gth{1}$ and $\gth{2}$ in this formalism is:
\begin{equation}
\begin{split}
\left( \gth{1} \circ \gth{1} \right) g =& \gth{1}\left(\gth{1}(g) \right) \SIM \gth{1}(g) \, ;\\
\left( \gth{2} \circ \gth{2} \right) g =& \gth{2}\left(\gth{2}(g) \right) \SIM \gth{2}(g) \, .\\
\end{split} 
\end{equation}
Taken together, it is possible to simplify all formulae wherever $\gth{1}$ and $\gth{2}$
appear next to each other more than once. The following examples should demonstrate how this works:
\begin{equation}
\begin{split}
\left( \gth{2} \circ \gth{3} \right) g &\equiv \left( \gth{2} \circ \gth{2} \circ \gth{1} \right) g 
\SIM \left( \gth{2} \circ \gth{1} \right) g
\equiv \gth{3} (g) \, ; 
\\
\left( \gth{1} \circ \gth{4} \right) g &\equiv \left( \gth{1} \circ \gth{1} \circ \gth{2} \right) g 
\SIM  \left( \gth{1} \circ \gth{2} \right) g
\equiv \gth{4} (g) \, , \\
\end{split}
\end{equation}
and in the same way
\begin{equation}
\begin{split}
 \gth{4} \circ \gth{3}  &\equiv  \gth{1} \circ \gth{2} \circ \gth{2} \circ \gth{1}  
\SIM  \gth{1} \circ \gth{2}  \circ \gth{1} 
\equiv \gth{1} \circ \gth{3}  \equiv \gth{4} \circ \gth{1} \, ; 
\\
\gth{3} \circ \gth{4}  &\equiv  \gth{2} \circ \gth{1} \circ \gth{1} \circ \gth{2}  
\SIM \gth{2} \circ \gth{1}  \circ \gth{2} \equiv \gth{2} \circ \gth{4}  \equiv \gth{3} \circ \gth{2} \, . \\
\end{split}
\end{equation}
These relationships can be used to structure the solution set generated starting from any particular perfect fluid sphere, and
moreover to classify which metrics can be produced by our theorems, and ones which cannot. 
For example, the idempotence property of theorem 1 and theorem 2 enables us to divide the class of perfect fluid spheres 
into seed metrics and non-seed metrics. Seed metrics can never be generated by using one of the two theorems $\gth{1}$ or $\gth{2}$,
while non-seed metrics are connected to other simpler metrics via one of these theorems.
We formalize this in the following subsection.

\subsection{Seed and non-seed metrics}

\noindent{\bf Definition (Seed metric):} 
Take a metric $g$ (or a parameterized class of metrics) and apply theorem 1 or theorem 2 on it.
Three different cases are possible:
\begin{itemize}
\item 
Each of the applications supplies us with a new solution. [$\gth{1}(g)\not\SIM g \not\SIM \gth{2}(g)$.] We define a metric with this behaviour as a seed metric.  (We shall soon see several examples of this behaviour.) 
\item 
Only one of the applications supplies us with a new solution, while the other one gives us the same metric we started with.  [$\gth{1}(g)\SIM g$ or $\gth{2}(g)\SIM g$.] These metrics are non-seed metrics.  (We shall soon see several examples of this behaviour.) 
\item 
Both applications give us the same metric we started with. [$\gth{1}(g)\SIM g \SIM \gth{2}(g)$.]  Metrics of this type are fixed points of the transformation process and we then also have $\gth{3}(g)\SIM g \SIM \gth{4}(g)$.
\end{itemize}
While we have encountered numerical examples that seem to exhibit this fixed point behaviour, we have no analytic proof for the existence of non-obvious fixed-point metrics. There is one obvious but not particularly useful example of a fixed point class of metrics. If we take the ODE in equation (\ref{ode_for_B}), and write down its most general solution as a functional of the arbitrary parameters $\zeta(r)$,  then any of our solution generating theorems applied to this most general solution will at most move us around in the parameter space characterizing the most general solution --- the most general solution of equation (\ref{ode_for_B}), or  equivalently equation (\ref{ode_for_zeta}), is thus an infinite-parameter fixed point of the generating theorems. But apart from this obvious example, it is unclear whether other fixed point classes of metric exist.\\

We can nevertheless develop several formal lemmata regarding fixed-point metrics.  For instance
\begin{lemma}
Suppose we have a metric such that $\forall \; \sigma$, $\epsilon$, $\lambda$
\begin{equation}
\gth{3}(\sigma,\epsilon,\lambda)\; g \SIM g,
\end{equation}
where we recall that $\SIM$ denotes equality up to redefinition of parameters.
Then in particular
\begin{equation}
\gth{1}(\lambda)\; g \SIM g \SIM \gth{2}(\sigma,\epsilon)\; g,
\end{equation}
and so
\begin{equation}
\gth{4}(\sigma,\epsilon,\lambda)\; g \SIM g.
\end{equation}

Conversely, suppose we have a metric such that
\begin{equation}
\gth{4}(\sigma,\epsilon,\lambda)\; g \SIM g,
\end{equation}
then
\begin{equation}
\gth{1}(\lambda)\; g \SIM g \SIM \gth{2}(\sigma,\epsilon)\; g,
\end{equation}
and so
\begin{equation}
\gth{3}(\sigma,\epsilon,\lambda)\; g \SIM g.
\end{equation}
\end{lemma}

\begin{proof} Trivial, note that $\gth{3}(\sigma,\epsilon,\lambda=0)=\gth{2}(\sigma,\epsilon)$ and $\gth{3}(\sigma=0,\epsilon=0,\lambda)=\gth{1}(\lambda)$. Then recall $\gth{4} = \gth{1}\circ\gth{2}$. Similarly for the converse.
\end{proof}

\begin{lemma}
Suppose we have a metric $g$ such that
\begin{equation}
\gth{4}\; g \SIM \gth{3}\; g
\end{equation}
and then define $g'$ by
\begin{equation}
\gth{4}\; g \SIM g' \SIM \gth{3}\; g.
\end{equation}
Then we have
\begin{equation}
\gth{4}\; g' \SIM g' \SIM \gth{3}\; g'
\end{equation}
so that $g'$ is a ``fixed point'' of both $\gth{3}$ and $\gth{4}$.
\end{lemma}

\begin{proof}
Note
\begin{equation}
\gth{4}\; g' \SIM \gth{4} \circ \gth{3}\; g \SIM \gth{1}\circ\gth{2}\circ\gth{1}\; g 
\SIM \gth{1}\circ\gth{3}\; g \SIM \gth{1}\circ\gth{4}\; g \SIM \gth{4}\; g = g'
\end{equation}
and similarly for $\gth{3}$:
\begin{equation}
\gth{3}\; g' \SIM \gth{3} \circ \gth{4}\; g \SIM \gth{2}\circ\gth{1}\circ\gth{2}\; g 
\SIM \gth{2}\circ\gth{4}\; g \SIM \gth{2}\circ\gth{3}\; g \SIM \gth{3}\; g = g'
\end{equation}
\end{proof}

Several other formal theorems along these lines can be constructed, but these seem the most important results.

Finally, among the formal properties enjoyed by the generating theorems, we mention the fact that theorems 3 and 4 are  ``conjugate'' to each other in the following sense
\begin{equation}
\gth{4} \equiv \gth{1}\circ\gth{2} =  \gth{1}\circ\gth{2} \circ \gth{1}\circ[\gth{1}]^{-1} =
 \gth{1}\circ\gth{3} \circ [\gth{1}]^{-1},
\end{equation}
and similarly (when the appropriate inverse $[\gth{2}]^{-1}$ exists)
\begin{equation}
\gth{3} \equiv \gth{2}\circ\gth{1} =  \gth{2}\circ\gth{1} \circ \gth{2}\circ[\gth{2}]^{-1} =
 \gth{2}\circ\gth{4} \circ [\gth{2}]^{-1}.
\end{equation}
We can write this as
\begin{equation}
\gth{3}(\sigma,\epsilon,\lambda)\sim\gth{4}(\sigma,\epsilon,\lambda),
\end{equation}
where $\sim$ denotes the concept of ``similarity'' under conjugation by invertible generating theorems. We wish to emphasise that similarity $\sim$ is a statement that holds for particular and fixed values of the parameters  $(\sigma,\epsilon,\lambda)$, as opposed to $\SIM$ which denotes equivalence under redefinition of parameters. If one is working numerically, it is much easier to ask questions involving similarity $\sim$. For analytic work, it is typically easier to ask questions involving equivalence $\SIM$.

\subsection{Two linking theorems}

The last two solution generating theorems we shall present are slightly different from those developed so far:
Consider a perfect fluid sphere solution $\left\{ \zeta_{0},B_{0}\right\}$ and extend it
to a new perfect fluid sphere $\left\{ \zeta_{0} Z_{0},B_{0} + \Delta \right\}$.  Previously, we had either set $Z_{0}=1$ and obtained theorem 1,
or we had set $\Delta=0$ and obtained theorem 2. In other words, we only changed one metric
component at a time.
(From this point of view theorem 3 and theorem 4 are, strictly speaking, not new theorems, in that they are replaceable by
iterations of theorem 1 and theorem 2 and vice versa.)
We now investigate what happens if we place no \emph{a priori} restrictions on $Z$ and $\Delta$, and allow both metric components to vary simultaneously. The differential equation
(\ref{ode_for_B}) for this problem now becomes
\begin{equation} \label{general_ode_for_B}
\left[ r (r \zeta_0 Z_{0} )' \right] \Delta'
+ \left[ 2 r^2 (\zeta_{0} Z_{0} )'' - 2 (r \zeta_{0} Z_{0})' \right] \Delta +
\left[ r^2 \zeta_0  B_0' + 4 r^2 \zeta_0' B_0 - 2 r \zeta_0 B_0\right] Z_0' + 2 r^2 \zeta_0 B_0 Z_0''  =0.
\end{equation}
Note that if $\Delta=0$ this becomes equation (\ref{ode_th1}), while if $Z_0=1$ this becomes equation (\ref{ode_th2}). 
In general, this ODE of first-order in $\Delta$, and --- as long $Z$ is not a constant --- inhomogeneous. In terms of
$\Delta$ this ODE can be solved explicitly and the result stated as a new theorem:
\begin{theorem}
Suppose $\left\{ \zeta_{0} , B_{0} \right\} $ describes a perfect fluid sphere, and let $Z_0(r)$ be arbitrary.
Define
\begin{eqnarray}
\Delta(\lambda,r) &=& \Delta_{0}(r)
\Bigg\{ \lambda - \int    
\frac{
\left[
\left(4r^2\zeta_{0}'B_{0}+r^2 \zeta_{0} B_{0}'-2r\zeta_{0}B_{0}\right) Z_{0}'
+ 2 r^2 \zeta_0 B_0 Z_0'' \right] 
\left\{ \zeta_0 + r \zeta_0' \right\}^2 
} { 
r^3 \;\left( r \zeta_{0} Z_{0}  \right)' \; \zeta_0^2\; Z_0^2 
}
\; 
\nonumber\\
&&
\qquad\qquad
\exp\left\{ -2 \int \frac{(\zeta_0 Z_{0})'}{ \zeta_0 Z_{0}} \; 
  \frac{ \zeta_0 Z_{0}- r\; (\zeta_0 Z_{0})' }{\zeta_0 Z_{0} + r  \;(\zeta_0 Z_{0})'} \; \d r\right\}
\d r    \Bigg\} ,
\end{eqnarray}
where
\begin{equation} 
\Delta_0(r)  =
 \left(    \frac{ \zeta_0 Z_{0}}{  \zeta_0 Z_{0} + r  \; (\zeta_0 Z_{0})'}\right)^2 \; r^2 \; 
\exp\left\{ 2 \int \frac{(\zeta_0 Z_{0})'}{ \zeta_0 Z_{0}} \; 
  \frac{ \zeta_0 Z_{0}- r\; (\zeta_0 Z_{0})' }{\zeta_0 Z_{0} + r  \;(\zeta_0 Z_{0})'} \; \d r\right\}.
\end{equation}
Then for all $\lambda$, the geometry defined by an arbitrary chosen $Z_0(r)$ and
setting
\begin{equation}
\d s^2 = - \zeta_0(r)^2 Z_{0}(r)^2 \; \d t^2 + {\d r^2\over B_0(r)+\Delta(\lambda,r) }
+ r^2 \d\Omega^2
\end{equation}
corresponds to a perfect fluid sphere. That is, the mapping
\begin{equation}
\gth{5}(\lambda): \left\{ \zeta_0 , B_0  \right\} \mapsto 
\left\{ \zeta_0 \; Z_0, B_0 + \Delta(\lambda,\zeta_0) \right\}
\end{equation}
takes perfect fluid spheres into perfect fluid spheres.
\end{theorem}

Note that if $Z_0(r)=1$ this simply reduces to theorem 1.
Re-arranging equation (\ref{general_ode_for_B}) in terms of $Z_{0}$ leads to
a second-order inhomogenous differential equation,  which cannot in general be solved for a prescribed $\Delta$, \emph{unless one imposes further constraints}. So further exploration in that direction is moot.
There is however a related theorem (which may be easier to understand) in terms of  parametric derivatives:
\begin{theorem}
Let $\{\zeta(\mu), B(\mu)\}$ denote a one-parameter class of perfect fluid spheres, so that the differential equation  (\ref{ode_for_B}) is satisfied for all $\mu$. Then
\begin{equation}
[r(r\zeta)'] \left({\d B\over\d\mu}\right)'+[2r^2\zeta''-2(r\zeta)']\left({\d B\over\d\mu}\right) 
+ 2 r^2 B \left({\d\zeta\over\d\mu}\right)'' + (r^2 B'-2rB)  \left({\d\zeta\over\d\mu}\right)' +(r B'-2B+2)  \left({\d\zeta\over\d\mu}\right)=0.
\end{equation}
In particular if $\d\zeta/\d \mu=0$ this reduces to the ODE (\ref{ode_th1}), while if $\d B/\d\mu=0$ this reduces to the 
ODE (\ref{ode_for_zeta}).
\end{theorem}
This is simply an alternative viewpoint on the previous theorem, emphasising the differential equation to be solved.

We again defer extensive discussion to the next section, but that this stage point out that if we invoke theorem 5 and apply it to Minkowski space, then making the choice $Z_0=1+r^2/a^2$ leads to the general Tolman IV metric --- that is:
\begin{equation}
\gth{5}(\hbox{Minkowski}; \; Z_0=1+r^2/a^2) = (\hbox{Tolman IV}).
\end{equation}
Even before we systematically start our classification efforts, it is clear that the solution generating theorems we have established will inter-relate \emph{many} of the standard perfect fluid spheres.

\subsection{Formal properties of the linking theorems}

Before turning to issues of systematic classification of perfect fluid metrics, we wish to establish a few formal properties of the linking theorems. To simplify the notation,  let us define the differential expression
\begin{equation}
D(\zeta,B) \equiv [r(r\zeta)']B'+[2r^2\zeta''-2(r\zeta)']B + 2\zeta = 
2 r^2 B \zeta'' + (r^2 B'-2rB) \zeta' +(r B'-2B+2)\zeta.
\end{equation}
Then the condition for a perfect fluid sphere is simply
\begin{equation}
D(\zeta,B)=0.
\end{equation}
Now define
\begin{equation}
D(\zeta_0,B_0+\Delta_0) = D(\zeta_0,B_0) + D_1(\zeta_0;\Delta_0).
\end{equation}
Then it is easy to check that
\begin{equation}
D_1(\zeta_0;\Delta_0) =  \left[ r (r \zeta_0 )' \right] \Delta_{0}'
+ \left[ 2 r^2 \zeta_{0}'' - 2 (r \zeta_{0})' \right] \Delta_{0}.
\end{equation}
The ODE for theorem 1, where we assume $\{\zeta_0,B_0\}$ is a perfect fluid sphere, is then
\begin{equation}
D_1(\zeta_0;\Delta_0)=0.
\end{equation}
Now define
\begin{equation}
D(\zeta_0 Z_0,B_0) = Z_0 D(\zeta_0,B_0) + D_2(\zeta_0,B_0;Z_0),
\end{equation}
then it is easy to check that
\begin{equation}
D_2(\zeta_0,B_0;Z_0) = 
(r^2 \zeta_0 B_0' + 4 r^2 \zeta_0' B_0 - 2 r \zeta_0 B_0) Z_0' + (2 r^2 \zeta_0 B_0) Z_0'' .
\end{equation}
The ODE for theorem 2, where we assume $\{\zeta_0,B_0\}$ is a perfect fluid sphere, is then
\begin{equation}
D_2(\zeta_0,B_0;Z_0)=0.
\end{equation}
Now let us consider any simultaneous shift in $\zeta$ and $B$, as considered in theorem 5. We have
\begin{equation}
D(\zeta_0 Z_0, B_0+\Delta_0) = D(\zeta_0 Z_0,B_0)+D_1(\zeta_0 Z_0;\Delta_0) = 
Z_0 D(\zeta_0,B_0) + D_2(\zeta_0,B_0;Z_0) + D_1(\zeta_0 Z_0;\Delta_0).
\end{equation}
But now let us write
\begin{equation}
D_1(\zeta_0 Z_0;\Delta_0) = Z_0 D_1(\zeta_0;\Delta_0) + D_{12}(\zeta_0;Z_0,\Delta_0),
\end{equation}
where a brief computation yields
\begin{equation}
D_{12}(\zeta_0;Z_0,\Delta_0) = r^2 \zeta_0 Z_0' \Delta_0' + [2r^2\zeta_0 Z_0''+4r^2\zeta_0'Z_0' - 2 r \zeta_0 Z_0' ] \Delta_0.
\end{equation}
Then all in all
\begin{equation}
D(\zeta_0 Z_0, B_0+\Delta_0) = Z_0 D(\zeta_0,B_0) + Z_0 D_1(\zeta_0;\Delta_0) +  D_{12}(\zeta_0;Z_0,\Delta_0) +  D_2(\zeta_0,B_0;Z_0).
\end{equation}
So if ${\zeta_0,B_0}$ and $\{\zeta_0 Z_0,B_0+\Delta_0\}$ are both perfect fluid spheres we must have
\begin{equation}
Z_0 D_1(\zeta_0;\Delta_0) +  D_{12}(\zeta_0;Z_0,\Delta_0) +  D_2(\zeta_0,B_0;Z_0)
= 0.
\end{equation}
 This is
 \begin{equation}
 Z_0\times(\hbox{ODE for theorem1}) + (\hbox{cross term}) + (\hbox{ODE for theorem2}) = 0.
\end{equation}
The cross term vanishes if either $Z_0=\hbox{constant}$ or $\Delta_0=0$ in which case we recover the usual theorem 1 and theorem 2. If we do things in the opposite order, then
 \begin{equation}
 D(\zeta_0 Z_0, B_0+\Delta_0) 
 = Z_0 D(\zeta_0,B_0+\Delta_0)+D_2(\zeta_0, B_0+\Delta_0;Z_0);
 \end{equation}
 \begin{equation}
 = Z_0 D(\zeta_0,B_0) + Z_0 D_1(\zeta_0,\Delta_0) + D_2(\zeta_0,B_0+\Delta_0;Z_0).
 \end{equation}
 We now have to compute
 \begin{equation}
 D_2(\zeta_0,B_0+\Delta_0;Z_0) =  D_2(\zeta_0,B_0;Z_0) + D_{21}(\zeta_0,B_0;Z_0,\Delta_0),
 \end{equation}
 and after a brief calculation
 \begin{equation}
 D_{21}(\zeta_0,B_0;Z_0,\Delta_0) = 2 r^2\zeta_0\Delta_0 Z_0'' +
  [r^2\zeta_0\Delta_0'+4 r^2\zeta_0'\Delta_0- 2 r \zeta_0\Delta_0] Z_0' = 
  D_{12}(\zeta_0,B_0;Z_0,\Delta_0).
 \end{equation}
 Thus the cross term is the same, no matter how you calculate it, and we still have the identity
\begin{equation}
D(\zeta_0 Z_0, B_0+\Delta_0) = Z_0 D(\zeta_0,B_0) + Z_0 D_1(\zeta_0;\Delta_0) +  D_{12}(\zeta_0;Z_0,\Delta_0) +  D_2(\zeta_0,B_0;Z_0).
\end{equation}
Provided  ${\zeta_0,B_0}$ and $\{\zeta_0 Z_0,B_0+\Delta_0\}$ are both perfect fluid spheres we again deduce
\begin{equation}
Z_0 D_1(\zeta_0;\Delta_0) +  D_{12}(\zeta_0;Z_0,\Delta_0) +  D_2(\zeta_0,B_0;Z_0)
= 0.
 \end{equation}
Now this gives us another way of looking at theorem 3 and theorem 4. For theorem 3 we first apply theorem 1 so we have the two equations
\begin{equation}
D_1(\zeta_0;\Delta_0) = 0;
\end{equation}
and
\begin{equation}
D_2(\zeta_0,B_0+\Delta_0;Z_0)\equiv D_{12}(\zeta_0;Z_0,\Delta_0) +  D_2(\zeta_0,B_0;Z_0) = 0.
 \end{equation}
Conversely, for theorem 4 where we first apply theorem 2 we see that we need to solve
\begin{equation}
D_2(\zeta_0,B_0;Z_0)= 0;
\end{equation}
and
\begin{equation}
D_1(\zeta_0 Z_0;\Delta_0) \equiv Z_0 D_1(\zeta_0;\Delta_0) +  D_{12}(\zeta_0;Z_0,\Delta_0) 
= 0.
 \end{equation}
 For theorem 5 we pick $Z_0$ arbitrarily, and solve the single ODE
 \begin{equation}
 D_1(\zeta_0 Z_0;\Delta_0)+ D_2(\zeta_0,B_0;Z_0)=0.
 \end{equation}
 This is a single first-order linear inhomogeneous ODE for $\Delta_0$, and hence solvable.
(In particular this makes it clear that theorem 5 is an \emph{inhomogeneous} version of theorem 1 with a carefully arranged ``source term'' $D_2(\zeta_0,B_0;Z_0)$. While theorem 5 is not ``idempotent'' it does satisfy the important formal property that:
\begin{lemma}
\begin{equation}
\gth{1}\circ\gth{5} \SIM \gth{5},
\end{equation}
which in particular tells us that the output from theorem 5 is never a seed metric.
\end{lemma}

\begin{proof}
Applying theorem 5 we need to solve
 \begin{equation}
 D_1(\zeta_0 Z_0;\Delta_0)+ D_2(\zeta_0,B_0;Z_0)=0,
 \end{equation}
 in order to map
 \begin{equation}
 \{\zeta_0,B_0\} \to \{\zeta_0Z_0,B_0+\Delta_0\}.
 \end{equation}
 Now apply theorem 1 to  $\{\zeta_0Z_0,B_0+\Delta_0\}$, this means we have to solve the homogeneous ODE
 \begin{equation}
D_1(\zeta_0 Z_0;\Delta_1)=0.
 \end{equation} 
But then, using properties of first-order ODEs
\begin{equation}
\left\{  D_1(\zeta_0 Z_0;\Delta_0)+ D_2(\zeta_0,B_0;Z_0)=0 \right\} \oplus
\left\{ D_1(\zeta_0 Z_0;\Delta_1)=0 \right\} 
\end{equation}
\begin{equation}
\implies
\left\{  D_1(\zeta_0 Z_0;\Delta_0+\Delta_1)+ D_2(\zeta_0,B_0;Z_0)=0 \right\}
\end{equation}
which is the ODE from theorem 5 back again.
\end{proof}
(The net result of this observation, as we shall see in the next section, is that theorem 5 can be used to connect one seed metric with the ``descendants'' of another seed metric.)

\section{Classifying  perfect fluid spheres}
\label{S:classify}
We will now see the power of these transformation theorems (solution generating theorems) by using them in a number of different ways:  to generate several new perfect fluid spheres, to relate various perfect previously known fluid spheres to each other,  and to classify the geometries we encounter.  First some minor comments and warnings: Despite comments made in~\cite{Delgaty},
Kuch~2~I $\equiv$ Tolman~V;   and Kuchb~I~b \emph{is} a perfect fluid for general values of its parameters. 
Furthermore RR--I $\equiv$ Einstein static; RR--V = Tolman V ($n=-5/4$). If we had not noted these degeneracies then our tables below would have been more extensive, but would have conveyed no extra information.


\begin{table}[!ht]
\centerline{Some selected perfect fluid solutions}
\bigskip
\begin{tabular}{|| l | l ||}
\hline
\hline
Name & Metric \\
\hline
\hline
Minkowski & $-\d t^2 + \d r^2 + r^2\d\Omega^2$ 
\\
Einstein static & $- \d t^2 + (1- {r^2}/{R^2})^{-1} \d r^2 + r^2 \d\Omega^2$
\\
de Sitter & $-(1 - {r^2}/{R^2})\, \d t^2 + (1 - {r^2}/{R^2})^{-1}\, \d r^2 + r^2 \d\Omega ^2$ 
\\
Schwarzchild Interior & $-\left(A - B \sqrt{1 - {r^2}/{R^2}}\right)^2 \d t^2 + (1- {r^2}/{R^2})^{-1} \d r^2 + r^2 \d\Omega^2$ 
\\
Schwarzchild Exterior & $-(1-2m/r)^2 \d t^2 + (1- 2m/r)^{-1} \d r^2 + r^2 \d\Omega^2$ 
\\
\hline
S1 & $- r^4 \d t^2 + \d r^2 + r^2 \d\Omega^2$
\\
K-O III & $- (A + Br^2)^2 \d t^2 + \d r^2 + r^2 \d\Omega^2$ 
\\
Kuch1 Ib & $ - (Ar + Br \ln r)^2 \d t^2+ 2 \d r^2 + r^2 \d\Omega^2$ 
\\
B--L & $- A (r^2/a^2)\d t^2 +  2(1+ r^2/a^2)^{-1} \d r^2 + r^2 \d \Omega^2$
\\
\hline
\vphantom{\Big|}Tolman IV &  
$- B^2 \left(1 + {r^2}/{a^2}\right) \d t^2 + 
\frac{1 + 2 {r^2}/{a^2}}{\left(1 - {r^2}/{b^2}\right)\left(1 + {r^2}/{a^2}\right)} \d r^2 
+ r^2 \d\Omega^2$ 
\\
\vphantom{\Big|}Tolman IV ($b\to\infty$)& 
 $- B^2 \left(1 + {r^2}/{a^2}\right) \d t^2 + \frac{1 + 2 {r^2}/{a^2}}{1 + {r^2}/{a^2}} \d r^2 + r^2 \d\Omega^2$ 
\\
\hline
Tolman V &  
$-B^2 r^{2(1+ n)}\d t^2 + (2-n^2)\; 
[1 - A r^{2(2-n^2)/(2+ n)}]^{-1} \d r^2 + r^2\d\Omega^2$
\\
Tolman V ($A\to0$) & $-B^2 r^{2(1+ n)}\d t^2 + (2-n^2)\d r^2 + r^2\d\Omega^2$
\\
Tolman VI & $-(Ar^{1-n} + Br^{1+n})^2 \d t^2 + (2-n^2)\d r^2 + r^2 \d\Omega^2$ 
\\
Tolman VII &
 $-B^2 \cos\left\{\ln\left[{\sqrt{1-2r^2/a^2+r^4/b^4}+r^2/a^2-b^2/a^2}\right]^{1/2}+\theta\right\}^2$ \\
 & \qquad 
 $+ (1-2r^2/a^2+r^4/b^4)^{-1} \d r^2 + r^2\d\Omega^2$
\\
Tolman VIII &
$ - A^2 r^{2(n-1)(n-4)/n}\left( 
{-n^2\over(n^2-4n+2)(n^2-8n+8)} + B r^{-(n^2-8n+8)/n} + C r^{-2(n^2-4n+2)/n} \right)
$
\\
\qquad 
&
$
+ \left( 
{-n^2\over(n^2-4n+2)(n^2-8n+8)} + B r^{-(n^2-8n+8)/n} + C r^{-2(n^2-4n+2)/n} \right)^{-1} 
\d r^2 + r^2\d\Omega^2$
\\
\hline
Kuch 68 II & $-\left(1-{2m/ r}\right)\d t^2 
+ \left[ \left(1-{2m/r}\right)\left(1 + C(2r-2m)^2 \right)\right]^{-1} \d r^2 + r^2\d\Omega^2$
\\
Kuch 68 I &
$ - \left(A\sqrt{1-2m/r}+
B\left[r^2/m^2+5r/m-30+15\sqrt{1-2m/r}\ln\{1-r/m-\sqrt{r(r-2m)}/m\}\right] \right)^2\d t^2$
\\
&\qquad
$+(1-2m/r)^{-1} \d r^2 + r^2 \d\Omega^2$
\\
M--W III & $-A r (r-a) +{7/4\over 1-r^2/a^2} \d r^2 + r^2 \d\Omega^2 $
\\
Kuch I b &
$ - r^2[A+B\,\arctanh(a/\sqrt{a^2+r^2})]^2\d t^2 + 2(1+r^2/a^2)^{-1}\d r^2 + r^2\d\Omega^2$
\\
Heint IIa (C=0) & $-(1+ar^2)^3 \d t^2 + [(1+ar^2)/(1-ar^2/2)] \d r^2 + r^2\d\Omega^2$
\\
Heint IIa &  $-(1+ar^2)^3 \d t^2 + \left[1-{3ar^2\over2(1+ar^2)} + {Cr^2\over(1+ar^2)\sqrt{1+4ar^2}} \right]^{-1} \d r^2 + r^2\d\Omega^2$
\\
\hline
\hline
\end{tabular}
\caption{Some well-known perfect fluid spheres and their corresponding metrics. Note that we have often reparameterized these metrics to make them easier to deal with, and so their appearance (but not the substance)  may differ from other sources~\cite{Delgaty,Skea,exact}.  Metric S1 is a special case of K--O III, Tolman V, and Tolman VI notable for its extreme simplicity.}
\end{table}

\begin{table}[!ht]
\centerline{Some apparently new perfect fluid solutions}
\bigskip
\begin{tabular}{|| l | l ||}
\hline
\hline
Name & Metric \\
\hline
\hline
Martin 1 & $ - ( Ar + Br \ln r)^2 \d t^2 + 2\frac{2A + 2B \ln r + B}{2A + 2B \ln r+B-Cr^2} \d r^2 + r^2 \d\Omega^2$ 
\\
Martin 2 & $- Ar(r-a) \d t^2 + {7\over 4} \left(1-{r^2\over a^2} - B {(r-a)r^{7/3}\over(4r-3a)^{4/3}} \right)^{-1}\d r^2 
+ r^2 \d\Omega^2$ 
\\
Martin 3 &$-(1 + ar^2)^2 \d t^2 + [1-br^2/(1+3ar^2)^{2/3}]^{-1} \d r^2 + r^2 \d\Omega^2$ 
\\
\hline

P1 & 
$-(1 + ar^2)^2 
\left[A + B\int {(1+ar^2)^{-2}\over\sqrt{1-br^2/(1+3ar^2)^{2/3}}}r \d r \right]^2 \d t^2 
+ [1-br^2/(1+3ar^2)^{2/3}]^{-1} \d r^2+ r^2 \d\Omega^2$ 
\\

P2 &  
$  - ( Ar + Br \ln r)^2 
\left[\sigma+\epsilon
\int  ( Ar + Br \ln r)^{-2} 
\left(2\frac{2A + 2B \ln r + B}{2A + 2B \ln r+B-Cr^2}\right)^{-1/2} r\d r \right]^2\d t^2 
$\\
& \qquad $
+ 2\frac{2A + 2B \ln r + B}{2A + 2B \ln r+B-Cr^2} \d r^2+ r^2 \d\Omega^2$
\\

P3 & $-(1+ar^2)^3 \left[ A + B {(5+2ar^2)\sqrt{1-ar^2/2}\over(1+ar^2)^{3/2}}\right]^2\d t^2 
+ [(1+ar^2)/(1-ar^2/2)] \d r^2 + r^2\d\Omega^2$
 \\
 
P4 & 
$- r^4\left( A + B\int {\d r\over r^3 \sqrt{1+\lambda r^{2/3}}}\right)^2 \d t^2 
+ (1+\lambda r^{2/3})^{-1} \d r^2 + r^2\d \Omega^2$\\
P4 & $- r^4\left( A + B\left[{15\over16}\lambda^3\tanh^{-1}(1/\sqrt{1+\lambda r^{2/3}}) 
- {1\over16}\sqrt{1+\lambda r^{2/3}} (8 r^{-2} - 10\lambda r^{-4/3} +15\lambda^2 r^{-2/3} ) \right]\right)^2 \d t^2 $
\\
 & \qquad $
+ (1+\lambda r^{2/3})^{-1} \d r^2 + r^2\d \Omega^2$\\

P5 & $- r \left(A\sqrt{r-a}+B\sqrt{r+a}\right)^2 
+{7/4\over 1-r^2/a^2} \d r^2 + r^2 \d\Omega^2 $
\\
P6 & $- r(r-a) \left(A+B\int{\d r\over(r-a)\sqrt{ 1-{r^2\over a^2} - 
B {(r-a)r^{7/3}\over(4r-3a)^{4/3}}  }}\right)^2 
+ {7\over4}\left(1-{r^2\over a^2}- 
B {(r-a)r^{7/3}\over(4r-3a)^{4/3}} \right)^{-1} \d r^2 + r^2 \d\Omega^2 $
\\

P7 &
$
- B^2 \left(1 + {r^2}/{a^2}\right) 
\left[ A + B \int {\sqrt{a^2-2r^2}\over\sqrt{b^2-r^2}(a^2-r^2)^{3/2}} r\d r \right]^2 \d t^2 + 
\frac{1 + 2 {r^2}/{a^2}}{\left(1 - {r^2}/{b^2}\right)\left(1 + {r^2}/{a^2}\right)} \d r^2 
+ r^2 \d\Omega^2$ 
\\

P8 &
$
-(1+ar^2)^3
 \left[ A + B  \int { r \; \d r \over (1+ar^2)^2 \sqrt{1-{3ar^2\over2(1+ar^2)} + {Cr^2\over(1+ar^2)\sqrt{1+4ar^2}}} }
 \right]^2 
 \d t^2 
+ \left[1-{3ar^2\over2(1+ar^2)} + {Cr^2\over(1+ar^2)\sqrt{1+4ar^2}} \right]^{-1} \d r^2
+ r^2 \d\Omega^2$ 
\\

\hline
\hline
\end{tabular}
\caption{Some apparently new perfect fluid spheres and their corresponding metrics. Sometimes the relevant integrals cannot be done in elementary form. When they can be done they are explicitly shown.}
\end{table}

Starting with the metric for any known perfect fluid sphere and successively applying theorem 1 and theorem 2 numerous times, will supply us with endless 
``new'' perfect fluid sphere solutions. Some of these ``new'' solutions might already be findable in the literature, some of them might be truly novel. Some of these solutions can be written down in fully explicit form. Some solutions are explicit but not elementary, in the sense that while the metric components can be exhibited as specific and explicit integrals, these integrals cannot be done in elementary form. Some solutions are so complex that present day symbolic manipulation programs quickly bog down.  (For specific symbolic computations we have used a vanilla installation of {\sf Maple}.)

\begin{table}[!ht]
\centerline{Some seed geometries and their descendants}
\bigskip

\begin{tabular}{|  l  |  l  |  l  |  l  | l |}
\hline
Seed & Theorem1 & Theorem2 & Theorem 3 & Theorem 4\\
\hline
Minkowski & Einstein static & K-O III & interior Schwarzschild & Martin 3\\ 
exterior Schwarzschild & Kuch68 II & Kuch 86 I  &    [integral] & [integral] \\
de Sitter& Tolman IV   & interior Schwarzschild  &  P7 & [integral] \\
Tolman V ($A=0$) & Tolman V &  Tolman VI & Wyman III & Wyman IIa\\
S1 & Tolman V ($n=+1$) & K--O III & P4 & Martin 3 \\
M--W III & Martin 2 &  P5 &  P6 & [integral] \\
Heint IIa (C=0) &  Heint IIa & P3 & P8 & [integral] \\
\hline
\hline
\end{tabular}
\caption{Seed solutions and their generalizations derived via theorems 1--4. The notation ``[integral]'' denotes a metric so complicated that explicitly writing out the relevant integral is so tedious that it does not seem worthwhile.}
\bigskip
\end{table}

\begin{table}[!ht]
\centerline{Some non-seed perfect fluid geometries and their descendants}
\bigskip
\begin{tabular}{|  l  |  l  |  l  |  l  | l |}
\hline
Base & Theorem1 & Theorem2 & Theorem 3 & Theorem 4\\
\hline
Tolman IV & Tolman IV &   P7  &   P7  & [integral] \\
B-L & B-L &Kuchb I b & Kuchb I b  & [integral]  \\
Heint IIa &  Heint IIa & P8 & P8 & [integral] \\
\hline

Tolman VI& Wyman IIa &  Tolman VI & [integral] & [integral] \\

Kuch1 Ib & Martin 1 & Kuch1 Ib &  P2 & Martin 1\\

K--O III & Martin 3&  K--O III &   P1  & Martin 3\\
\hline



\hline
\end{tabular}
\caption{Non-seed solutions and their generalizations.}
\bigskip

\end{table}

To summarize the situation we present several tables and diagrams.  Two tables are used to provide the names and explicit metrics for the perfect fluid spheres we consider. Two other tables are used to describe the inter-relationships of these perfect fluid spheres under $\gth{1}$, $\gth{2}$, $\gth{3}$, and $\gth{4}$.  
In these tables the notation  ``[integral]'' means that it is definitely a novel perfect fluid solution but the metric components involve an explicit integral that does not appear to be do-able by elementary methods, and is so complicated that it does not seem worthwhile to even write it down.
Recall that a seed metric is one for which theorem 1 and theorem 2 both yield metrics  distinct from the seed: $\gth{1}(g)\not\SIM g\not\SIM\gth{2}(g)$. In contrast for non-seed metrics one or the other of these theorems is trivial, either $\gth{1}(g)\SIM g$ or $\gth{2}(g)\SIM g$.

By considering the idempotence of theorem 1 and theorem 2, and the fact that theorem 3 and theorem 4 can be expressed in terms of the first two theorems, it is possible to structure, and therefore to graphically visualize the relationship between all metrics generated from a given seed metric.
We demonstrate this behaviour starting with Minkowski spacetime as seed metric,  checking if it is possible to create endless new perfect fluid spheres  starting with this  trivial seed. The first few steps can be carried out explicitly, and show that Minkowski space generates several well known interesting perfect fluid models.

\begin{figure}[!ht]
\centering
\includegraphics[width = \textwidth]{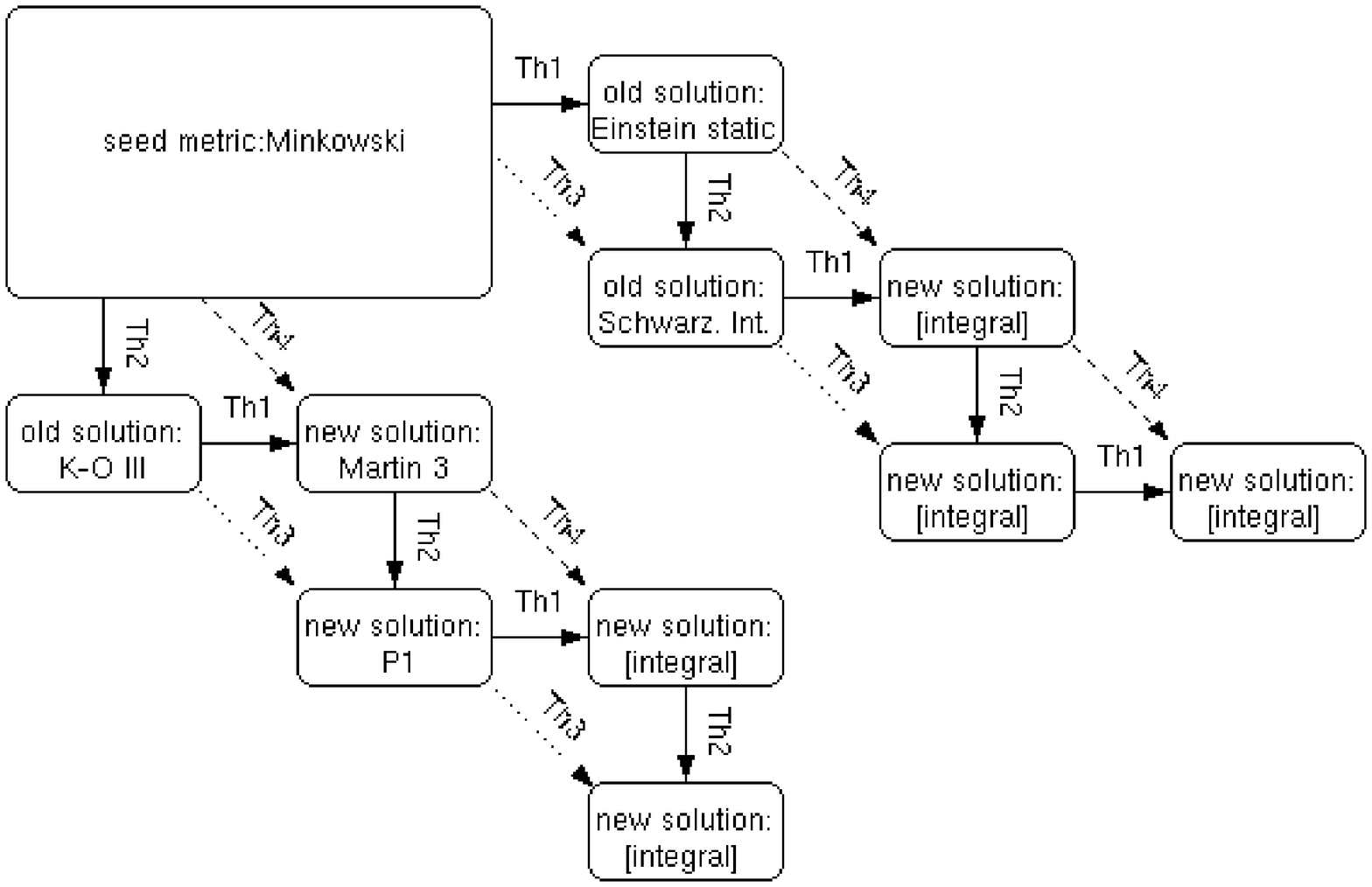}
\label{structure}
\caption{\label{Structure_Graph1}Structure graph for Minkowski space as seed metric.}
\end{figure}

In figure \ref{Structure_Graph1} all the ideas from the previous sections are used.
Each box represents a specific metric (perfect fluid sphere) while the arrows correspond to the 
application of the different theorems. The horizontal arrows correspond 
to an application of theorem 1 and the vertical arrows to an application 
of theorem 2. In addition, the vectors pointing along the diagonals can 
either indicate an application of theorem 3 or theorem 4. A dotted arrow 
corresponds to the application of theorem 3 while the dashed arrow 
represents an application of theorem 4.

Figure \ref{Structure_Graph1} shows that after applying theorem 1 to the Minkowski seed metric , we get the Einstein static universe.  By the idempotence of theorem 1, 
$n$ applications of  $\gth{1}$ to the seed metric still results in the Einstein 
static. Similary, any number of  applications of theorem 2 after the first (which leads to the K--O III solution) does not give us any further new solutions (see figure \ref{Idempotence_Theorem2}).
We can also see that the first and second generating theorems are not 
commutative.  Application of theorem 1 and theorem 2 in that order to the Minkowski seed 
metric results in the Schwarzchild Interior geometry, whereas application 
of theorem 1 after theorem 2 gives us the Martin 3 solution, and the three-parameter Schwarzschild interior solution is quite distinct from the three-parameter Martin 3 solution.

\begin{figure}[!htb]
\includegraphics[width = \textwidth]{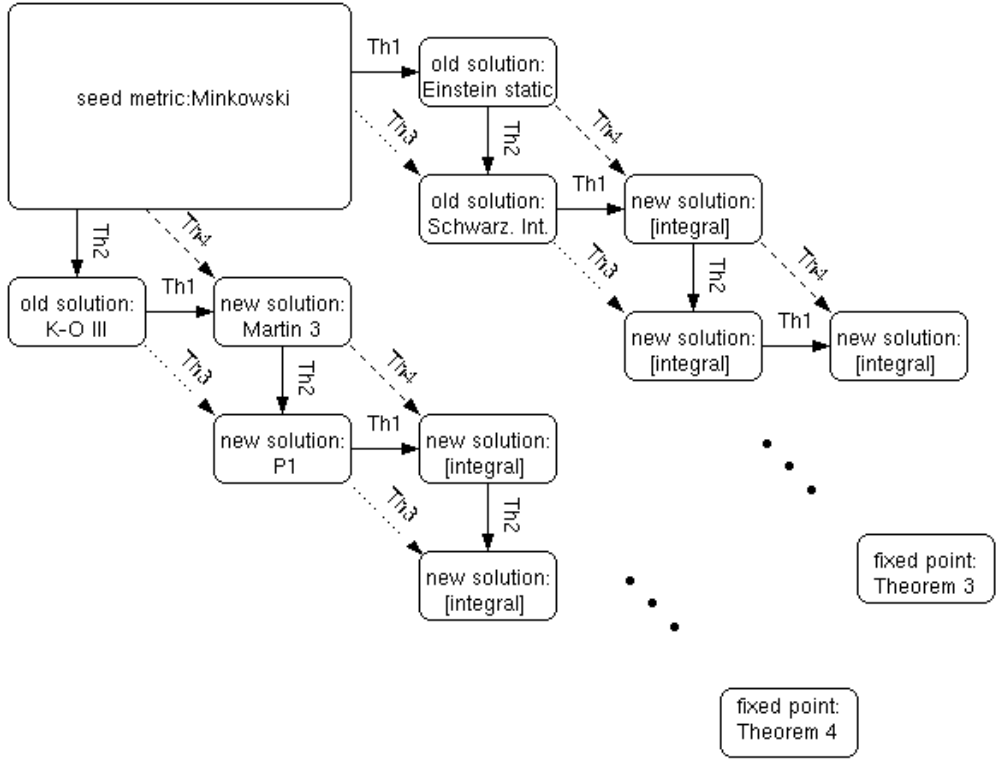}
\caption{\label{Structure_Graph2} Numerical implementations of Theorem 3 and Theorem 4 can be used to investigate their behaviour for $n$ applications, as the number of applications becomes large.
For specific choices of parameters $\lambda_{i}$, $\sigma_{i}$ and $\epsilon_{i}$ successive applications of Theorem 3 and Theorem 4 appear to be approaching a limit, in the sense that both metric components seem to  converge toward  fixed points.}
\end{figure}

Indeed, it seems as if it is possible to create endless new solutions for a perfect fluid sphere out
of the Minkwoski metric (or any other of the seed metrics). 
After several iterations the calculations become more complex, and  can no longer be carried out analytically. We then resort to numeric computation  to 
find out whether theorem 3 and theorem 4 have some sort of numerical limit, a numerical fixed point, or not. 
Depending on the choices made for the parameters $(\lambda,\sigma,\epsilon)$ both theorems converge very quickly. Specifically, we used {\sf Matlab} to numerically analyze the evolution of  $\zeta(r)$ and $B(r)$ after applying theorem 3 several times to the Minkowski metric. 
Numerical results are summarized in figure \ref{Structure_Graph2}. This figure indicates that theorem 3 and theorem 4 both appear to have well  defined numerical limits, though we have no direct analytic solution for the fixed point metric.

\section{Discussion}
Using Schwarzschild coordinates we have developed several transformation theorems that map perfect fluid spheres into perfect fluid spheres, and have used these transformations as a basis for classifying different types of perfect fluid sphere solutions. 
If we apply these theorems on a known perfect fluid sphere, different
solutions are often obtained. While some of the solutions are already known in the literature, most of them are novel.
Therefore, we have developed a tool to generate new solutions for a perfect fluid sphere, which does not
require us to directly solve the Einstein equations. (Of course the whole procedure was set up in such a way that we are implicitly and indirectly solving the Einstein equations, but the utility of the transformation theorems is that one does not have to go back to first principles for each new calculation.)

In addition, we have also established several relationships among the generating theorems. 
Previously, all metrics seemed to have nothing more in common than representing a
perfect fluid sphere. 
We have developed the concept of a seed metric, which is one that cannot be generated by our theorems.
Starting with a seed metric and applying our theorems, it is possible to structure, therefore to visualize, the relationship between all metrics generated from a given seed metric in a graph. Based on this example it is possible to create endless new solutions out of the Minkowski metric. We also used a numeric program to investigate whether theorem 3 and theorem 4 have fixed point limits. Both seem to converge very quickly.

In summary, the class of static perfect fluid sphere solutions to the Einstein equations exhibits a deep and somewhat unexpected structure --- the transformation theorems of this article provide a new and distinct way of looking at the problem, and yield a new way of viewing the inter-relationships between different static fluid spheres.



\end{document}